\documentclass[paper]{pasj00}
\Received{17 February, 2014 }
\Accepted{25 March, 2014}
\Published{$\langle$publication date$\rangle$}
\SetRunningHead{Large-scale Dynamo in Rotating Stratified Convection}{Masada and Sano}

\usepackage{times}
\usepackage{graphicx}
\usepackage{bm}

\begin{document}

\title{Long-term Evolution of Large-scale Magnetic Fields \\in Rotating Stratified Convection}
\author{Youhei \textsc{Masada}\altaffilmark{1} and Takayoshi \textsc{Sano} \altaffilmark{2} }
%\thanks{Last update: January 19, 2007}}
\altaffiltext{1}{%
   Department of Computational Science, Kobe University; \\
   Kobe 857--8501; ymasada@harbor.kobe-u.ac.jp}
%   \email{ymasada@harbor.kobe-u.ac.jp}
\altaffiltext{2}{%
   Institute of Laser Engineering, Osaka University; \\
   Suita, Osaka 565--0871;sano@ile.osaka-u.ac.jp}


\maketitle

%%%%%%%%%%%%%%%%%%%%%%%%%%%%%%%%%%%%%%%%%%%%%%%%%%%%
\begin{abstract}
Convective dynamo simulations are performed in local Cartesian geometry. We report the first successful 
simulation of a large-scale oscillatory dynamo in rigidly rotating convection without stably stratified layers. 
A key requirement for exciting the large-scale dynamo is a sufficiently long integration time comparable to the ohmic diffusion time. 
By comparing two models with and without stably stratified layers, their effect on the large-scale dynamo is also studied. 
The spatiotemporal evolution of the large-scale magnetic field is similar in both models. However, it is intriguing that the magnetic 
cycle is much shorter in the model without the stable layer than with the stable layer. This suggests that the stable layer 
impedes the cyclic variations of the large-scale magnetic field.  
\end{abstract}
%%%%%%%%%%%%%%%%%%%%%%%%%%%%%%%%%%%%%%%%%%%%%%%%%%%%
\section{Introduction}
A grand challenge in astrophysics is to understand a self-organizing property of magnetic fields in highly 
turbulent flows. The solar magnetism is the front line of this area. The solar magnetic field shows a remarkable 
spatiotemporal coherence even though it is generated by turbulent convection operating within its interior. 
Our understanding on the solar magnetism has been accelerated over the past decade in response to the 
broadening, deepening and refining of numerical dynamo models \citep{charbonneau10,brandenburg+12,miesch12}. 
However, it is still unclear what dynamo mode is excited in the solar interior and how it regulates the magnetic cycle. 

Various geometries have been applied to the numerical dynamo modeling; global spherical shell geometry 
(e.g., \cite{gilman+81,brun+04,ghizaru+10,masada+13}), spherical-wedge geometry (e.g., \cite{brandenburg+07,kapyla+10}), 
and local Cartesian geometry (e.g., \cite{cattaneo+91, brandenburg+96}). Among them, the local Cartesian geometry is 
the most simplified one and is often used for distilling the physical essence of the convective dynamo process by 
more accurately resolving convective eddies.

A long-standing goal in the numerical dynamo modeling in the local Cartesian geometry is to realize the successful simulation 
of self-organized and self-sustained large-scale magnetic fields, so-called ``large-scale dynamos", by rotating convection 
alone without mean shear flow. The mean-field dynamo theory predicts that the rigidly rotating convection can generate net 
helicity and then excite the large-scale dynamo even without the mean shear effect via a stochastic process, which is known 
as the $\alpha$-effect  (\cite{moffatt+78,krause+80}). However, no evidence of the large-scale dynamo was found in earlier studies of 
the rigidly rotating convection (e.g., \cite{cattaneo+06,tobias+08}).

\citet{kapyla+09} brought a breakthrough in the dynamo modeling in the local system. 
They were the first to demonstrate that the rigidly rotating convection can excite the large-scale dynamo in the local system that consists 
of the convection layer and the stably stratified layers. Subsequently, the oscillatory behavior of the large-scale magnetic field was 
reported in \citet{kapyla+13}. Since the large-scale dynamo can be excited only when the Coriolis number is large, they concluded that 
the absence of the large-scale dynamo in earlier studies is caused by the slow rotation speed. 

However, even in the sufficiently-rapid rotating convection, Favier \& Bushby (2013) could not find the evidence for the large-scale dynamo 
in the local system with the convection zone alone. They suggested that the essential part for the large-scale dynamo might be the stably 
stratified layer assumed in the model of \citet{kapyla+09} rather than the rapid rotation. 

Therefore, at present, the key requirement for the large-scale dynamo is still controversial. The purpose of this work is to find the 
evidence of the large-scale magnetic field in the system only with the convection zone in order to demonstrate that the rigidly rotating 
convection is a sufficient condition for the large-scale dynamo. In addition, by comparing two convective dynamo models with and without stably 
stratified layers, we will discuss their effect on the large-scale dynamo. 

%%%%%%%%%%%%%%%%%%%%%%%%%%%%%%%%%%%%%%%%%%%%%%%%%%%%
\section{Model Setup}
%%%%%%%%%%%%%%%%%%%%%%%%%%%%%%%%%%%%%%%%%%%%%%%%%%%%
%%%%%%%%%%%%%%%%%%%%%%%%%%%%%%%%%%%%%%%%%%%%%%%%%%%%%%%%%%%%%%%%%%%%%%%%
\begin{figure}[tbp]
\begin{center}
\includegraphics[width=85mm]{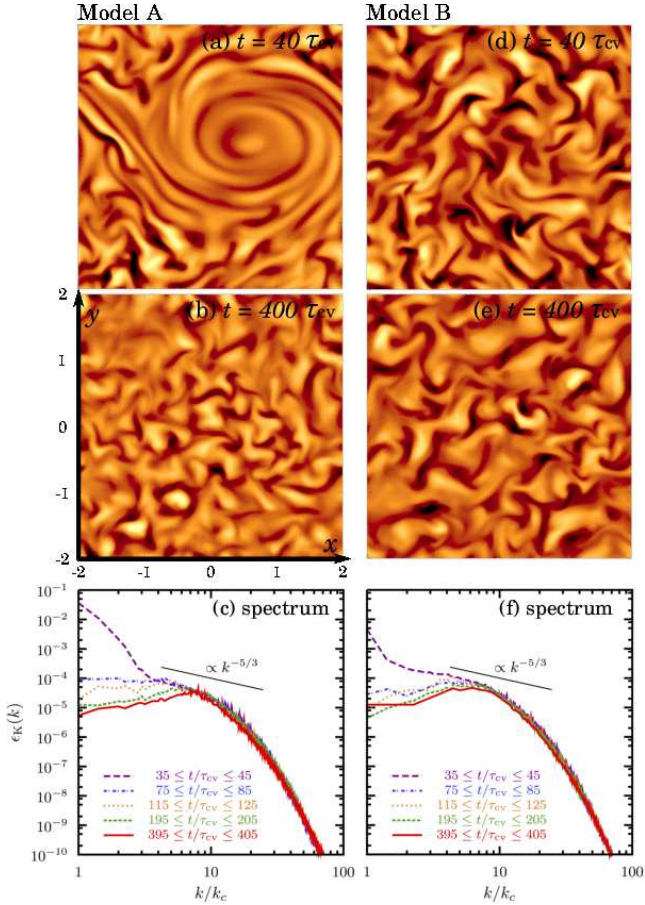}
\end{center}
\caption{Distribution of the radial velocity on the horizontal plane at the middle of the convection zone when (a) $t=40\tau_{\rm cv}$ 
and (b) $t=400\tau_{\rm cv}$ for the model A (panels (d) and (e) are those for the model B). Panels (c) and (f) are kinetic energy spectrum 
for the models A and B. Power-law slopes proportional to $k^{-5/3}$ are shown for reference.}\label{fig0}
\end{figure}
%%%%%%%%%%%%%%%%%%%%%%%%%%%%%%%%%%%%%%%%%%%%%%%%%%%%%%%%%%%%%%%%%%%%%%%%
We numerically solve two convective dynamo systems in local Cartesian domain:  one-layer system (Model A) only 
with convection zone of thickness $d$ ($z_1 \le z < z_2$), and three-layers system (Model B) consisting of upper 
isothermal cooling layer of depth $0.15d$ ($z_0 \le z \le z_1$), middle convection layer of depth $d$ ($z_1 \le z < z_2$) and 
bottom stably stratified layer of depth $0.85d$ ($z_2 \le z < z_3$), where  the $x$- and $y$-axes are taken to be 
horizontal and $z$-axis is pointing downward. The aspect ratio between the thickness of the convection layer and 
the box width ($W$) sets to be $W/d=4$ for both models. The setups in the models A and B are similar to those used in 
\citet{favier+13} and \citet{kapyla+09}, respectively. 

The basic equations are the compressible MHD equations in the rotating frame of reference with a constant angular velocity 
${\bm \Omega} = -\Omega_0 {\bm e}_z$ 
\begin{eqnarray}
\frac{\partial \rho}{\partial t} & = & - \nabla\cdot (\rho {\bm u})  \;, \label{eq1} \\ 
\frac{{D }{\bm u}}{ D t}  & = & - \frac{\nabla P}{\rho}  
+ \frac{{\bm J}\times{\bm B} }{\rho} - 2{\bm \Omega}\times{\bm u} + \frac{\nabla \cdot {\bm \Pi}}{\rho} + {\bm g}  \;, \ \ \ \ \label{eq2} \\
\frac{D\epsilon }{D t} & = & - \frac{P\nabla\cdot {\bm u}}{\rho}  + \mathcal{Q}_{\rm heat} - \frac{\epsilon - \epsilon_0}{\tau (z)} \;, \label{eq3} \\
\frac{\partial {\bm B} }{\partial t} & = & \nabla \times ({\bm u} \times {\bm B} - \eta_0 {\bm J})\;, \label{eq4}
\end{eqnarray}
where ${\bm J} = \nabla \times {\bm B}/\mu_0$ is the current density, ${\bm g} = g_0 {\bm e}_z$ is the gravity, $\epsilon$ is the specific 
internal energy. The viscosity, magnetic diffusivity, and thermal conductivity are represented by $\nu_0$, $\eta_0$, and $\kappa_0$, respectively. The last 
term in equation (3) works only in the model B and describes a cooling at the top of the domain with the cooling time $\tau(z)$ which has a smooth profile 
connecting to the convection layer, where $\tau(z_1) = \infty $. 

The viscous stress ${\bm \Pi}$ is written by ${\bm \Pi}  =   2\rho \nu_0 {\bm S}$ with the strain rate tensor
 \begin{equation}
S_{ij}  =  \frac{1}{2}\left( \frac{\partial u_i}{\partial x_j} +  \frac{\partial u_j}{\partial x_i} - \frac{2}{3}\delta_{ij}\frac{\partial u_i}{\partial x_i} \right) \;.  \label{eq5}
\end{equation}
The heating term $\mathcal{Q}_{\rm heat}$ consists of the thermal conduction, viscous heating and Joule heating, 
\begin{equation}
\mathcal{Q}_{\rm heat} =   \frac{\nabla\cdot(\kappa_0 \nabla\epsilon)}{\rho} + 2\nu_0 \bm{S}^2 + \frac{\mu_0\eta_0\bm{J}^2}{\rho}\;. \label{eq6}
\end{equation}
We assume a perfect gas law $P = (\gamma -1)\rho \epsilon$ with $\gamma = 5/3$. 

The initial hydrostatic balance is described by a piecewise polytropic distribution with the polytropic index $m$,
\begin{equation}
\frac{{\rm d} \epsilon}{{\rm d} z} = \frac{g_0}{(\gamma-1)(m + 1)} \;.
\end{equation}
We choose $m=1$ for the convection layer, and $m=3$ for the stable layer. The thermal conductivity is 
determined by requiring a constant vertical heat flux throughout the domain. 

Normalization quantities are defined by setting $d=g_0=\rho_0=\mu_0=1$, where $\rho_0$ is the initial density at $z=z_0$. 
The velocity normalization corresponds to $\sqrt{d g_0} = 1$. 
The stratification level is controlled by the normalized pressure scale height at the 
surface defined by $\xi = H_p/d = (\gamma-1)\epsilon_0/(g_0d)$, where $\epsilon_0$ is the specific internal energy at 
$z=z_0$. In this work, we use $\xi = 0.3$, yielding a density contrast between top and bottom of the convection zone about $5$. 

We define the Prandtl, magnetic Prandtl, and Rayleigh numbers by
\begin{equation}
{\rm Pr}  =  \frac{\nu_0}{\chi_0} \;,\ {\rm Pm} = \frac{\nu_0}{\eta_0}\;, \ {\rm Ra}  =  \frac{g_0 d^4}{\chi_0 \nu_0}\left[ \frac{\nabla - \nabla_{\rm ad}}{H_{p}} \right] \;, 
\end{equation}
where $\chi_0 \equiv \kappa_0/\gamma\rho$ is the thermal diffusivity, $\nabla - \nabla_{\rm ad}$ is the superadiabatic temperature 
gradient with $\nabla_{\rm ad} = 1-1/\gamma$, $\nabla = (\partial \ln T/\partial \ln P)$, and $H_{p}$ is the pressure scale height.
%Note that  
The variables ($\rho$, $\nabla$, and $H_p$) in equation (8) are evaluated at the mid-convection zone of the depth $z_m = (z_2-z_1)/2$. 

In the following, the volume average in the convection zone and the horizontal average are denoted by single angular 
brackets with subscript ``\rm v" and subscript ``\rm h", respectively. The time-average of each spatial mean is denoted by additional angular brackets. 
The relative importance of rotation to the convection is measured by the Coriolis number ${\rm Co} = 2\Omega_0d/u_{\rm cv}$, 
where $u_{\rm cv} \equiv \sqrt{\langle \langle u_z^2 \rangle\rangle_{\rm v}}$ is the mean convective velocity. 
The convective turn-over time and the %volume-averaged 
equipartition field strength are defined, respectively, by $\tau_{\rm cv} \equiv d/u_{\rm cv}$ and 
$B_{\rm eq} \equiv \sqrt{\langle \langle \mu_0 \rho {\bm u}^2 \rangle\rangle_{\rm v}} $.

In the horizontal directions, all the variables are assumed to be periodic. Stress-free boundary conditions are used in the vertical 
direction for the velocity. Perfect conductor and vertical field boundary conditions are used for the magnetic field at the bottom  
and top boundaries, respectively. A constant energy flux which drives the convection is imposed on the bottom boundary. 
The internal energy is fixed on the top boundary. 

The equations (1)--(6) are solved by the second-order Godunov-type finite-difference scheme which employs 
an approximate MHD Riemann solver developed by \citet{sano+98}. The magnetic field evolution is calculated with CMoC-CT method \citep{clarke96}. 
Non-dimensional parameters $Pr=1.4$, $Pm=4.0$, $Ra = 3.9\times10^6$, constant angular velocity of $\Omega_0 = 0.4$ and the same grid spacing 
are adopted for both models.  
The total grid size is $256$ (in $x$) $\times 256$ (in $y$) $\times$ $64$ (in $z$) for the model A, 
and $256$ (in $x$) $\times 256$ (in $y$) $\times$ $128$ (in $z$) for the model B. A small random perturbation is added to the velocity and magnetic fields 
when the calculation starts. 
%%%%%%%%%%%%%%%%%%%%%%%%%%%%%%%%%%%%%%%%%%%%%%%%%%%%%%%%%%%%%%%%%%%%%%%%
\begin{figure*}[tbp]
\begin{center}
\includegraphics[width=150mm]{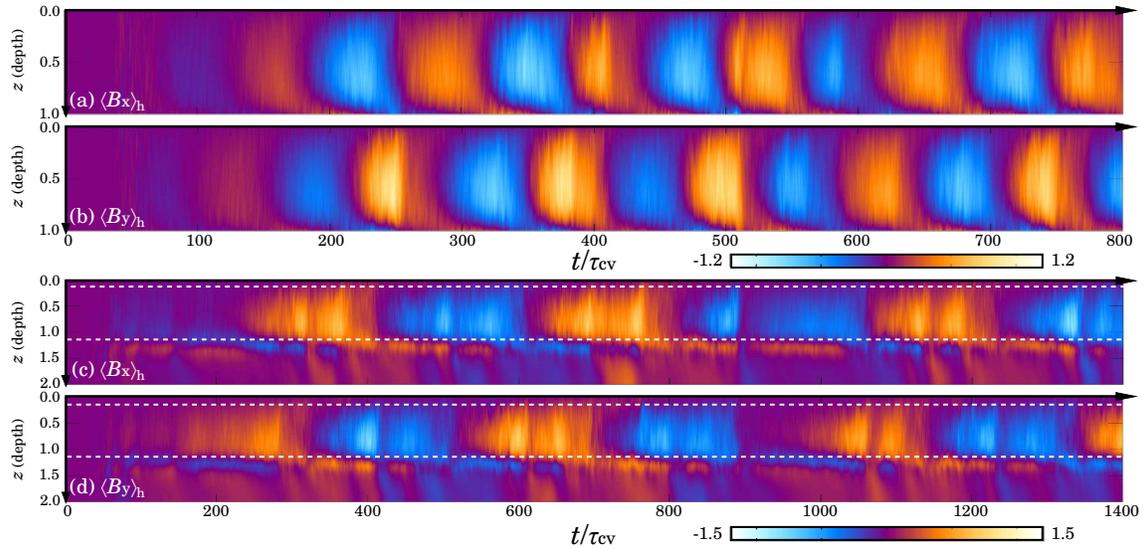}
\end{center}
\caption{Time-depth diagram of the horizontally-averaged horizontal magnetic field. Panels (a) and (b) [(c) and (d)] demonstrate $\langle B_x \rangle_{\rm h}$ and $\langle B_y \rangle_{\rm h}$ normalized by $B_{\rm eq}$ for the model A [model B]. The orange and blue tones denote the positive 
and negative strengths of the magnetic field. }
\label{fig2}
\end{figure*}
%%%%%%%%%%%%%%%%%%%%%%%%%%%%%%%%%%%%%%%%%%%%%%%%%%%%%%%%%%%%%%%%%%%%%%%%
%%%%%%%%%%%%%%%%%%%%%%%%%%%%%%%%%%%%%%%%%%%%%%%%%%%%%%%%%%%%%%%%%%%%%%%%
\section{Simulation Results}
%%%%%%%%%%%%%%%%%%%%%%%%%%%%%%%%%%%%%%%%%%%%%%%%%%%%%%%%%%%%%%%%%%%%%%%%
\subsection{Properties of Convective Dynamo}
After the convective motion sets in, the system reaches a saturated state at $t \simeq 250\tau_{\rm cv}$ for both models. The mean convective 
velocity is evaluated there as $u_{\rm cv} = 0.017$ $(0.019)$, providing $B_{\rm eq} = 0.045$ $(0.045)$, $\rm{Co} = 47$ $(42)$ and 
$\tau_{\rm cv} = 58.8$ $(52.6)$ for the model A (model B). Since a sufficient scale separation between the convective eddies and the box scale 
is known as a necessary ingredient for the large-scale dynamo (e.g., \cite{brandenburg+05,kapyla+09}), we have chosen the relatively rapid 
rotation ($\rm{Co} \gtrsim 40$), yielding small convective cells relative to the box scale.

Shown in Figure 1 is the distribution of the radial velocity in the horizontal plane at $z=z_m$ when (a) $t=40\tau_{\rm cv}$ 
and (b) $t=400\tau_{\rm cv}$ for the model A [panels (d) and (e) are those for the model B]. The darker and lighter tones depict downflow and 
upflow velocities. In Figure 1(c) and (f), the temporal evolutions of two-dimensional kinetic energy spectra for the models A and B are shown. 
Note that a two-dimensional Fourier spectrum of the kinetic energy at the each depth is projected onto a one-dimensional 
wavenumber $k^2 = k_x^2 + k_y^2$ and then is averaged over the convection zone and the time span shown in the figure legend. 
The different lines correspond to different time spans. The horizontal axis is normalized by $k_c = 2\pi/W$. 

In the kinematic phase of the model A (left top), we can find the appearance of large-scale vortices, which have been discovered 
in earlier studies of rotating convections (e.g., \cite{chan07}). In this phase, the spectrum has a peak at %the small wavenumber 
$k/k_c = 1$ (purple long-dashed line). As found by \citet{kapyla+13}, the large-scale vortices decay as the magnetic fields grow, and finally disappears at the 
saturated phase. In the saturated phase, the convective motion is characterized by the broader and slower upflow surrounded by 
narrower and faster downflow lanes. The spectrum has a peak at around $k/k_c \simeq 6$ for both models. 
Since there is no physical mechanism for the symmetry breaking in the horizontal directions, mean horizontal shear flow is 
absent in our simulations. In contrast, a mean kinetic helicity naturally arises from the up-down asymmetry in the convective motion 
(e.g., \cite{spruit+90}). This could play a prominent role in sustaining large-scale dynamo in our systems. 

Figure 2 depicts the time-depth diagram of horizontally-averaged horizontal magnetic fields. Shown in panels (a) and (b) [(c) and (d)] are 
$\langle B_x \rangle_{\rm h} (t,z)$ and $\langle B_y \rangle_{\rm h} (t,z)$ normalized by $B_{\rm eq}$ for the model A [model B]. The orange and blue tones 
denote the positive and negative strengths of the magnetic field. The region between white dashed lines in panels (c) and (d) corresponds to the convection 
layer. In the saturated phase of $t \gtrsim 250\tau_{\rm cv}$, the large-scale magnetic field is spontaneously organized in the bulk of the convection zone 
for both models. The strength of the large-scale field maximally exceeds $B_{\rm eq}$ at the mid convection zone, and averagely is about $20 \%$ of the 
total field strength. %In contrast, the vertical field does not show any coherent signatures, and is fully dominated by fluctuating component, i.e., 
%$\langle B_z \rangle_{\rm h} = 0$. 

The large-scale magnetic field shows a well-regulated oscillatory behavior. The strong magnetic field appears at the middle of the convection zone 
and propagates from there to top and base of the convection zone for both models. It is intriguing that the oscillation period of the large-scale 
magnetic field is shorter in the model A than in the model B. The polarity is reversed with the period of about $70 \tau_{\rm cv}$ for 
the model A and $200 \tau_{\rm cv}$ for the model B. It is noteworthy that there is a phase difference of about $\pi/2$ between 
$\langle B_x \rangle_{\rm h} $ and $\langle B_y \rangle_{\rm h}$. %The observed oscillatory behavior is similar to that seen in the simulation of \citet{kapyla+13}.
The observed oscillatory behavior in the model B is similar to that seen in \citet{kapyla+13}, but its cycle period is about twice as long as that in their model.
%is reminiscent of the solar butterfly diagram although 
%there is a difference in the propagation direction between the simulated field and the sunspot field. 
%%%%%%%%%%%%%%%%%%%%%%%%%%%%%%%%%%%%%%%%%%%%%%%%%%%%%%%%%%%%%%%%%%%%%%%%
\begin{figure*}[htbp]
\begin{center}
\includegraphics[width=165mm]{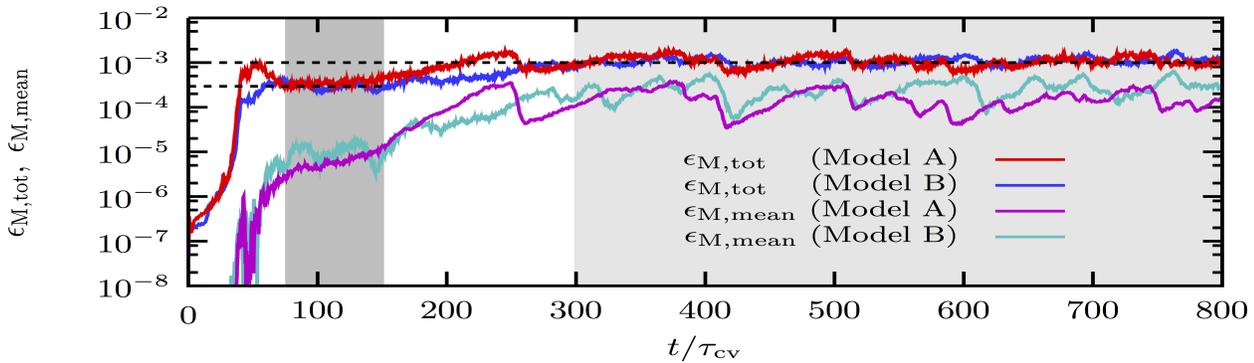}
\end{center}
\caption{Time-evolutions of the volume-averaged energy of the total magnetic field 
$\epsilon_{\rm M,tot} \equiv \langle {\bm B}^2\rangle_{\rm v}/2\mu_0$ (red line) and that of the mean-magnetic component defined by 
$\epsilon_{\rm M, mean} \equiv \langle {\bm B}\rangle_{\rm v}^2/2\mu_0$ (purple line) for the model A. The blue and cyan lines are 
those for the model B. }
\label{fig3}
\end{figure*}
%%%%%%%%%%%%%%%%%%%%%%%%%%%%%%%%%%%%%%%%%%%%%%%%%%%%%%%%%%%%%%%%%%%%%%%%
\subsection{Two-Step Saturation Process of the Magnetic Field}
%%%%%%%%%%%%%%%%%%%%%%%%%%%%%%%%%%%%%%%%%%%%%%%%%%%%%%%%%%%%%%%%%%%%%%%%
In earlier studies of rigidly rotating convections, the large-scale dynamo was not found in the system only with the convection zone 
\citep{cattaneo+06,favier+13}. Favier \& Bushby (2013) suggested that the stably stratified layer, which was considered in \citet{kapyla+09}, 
might play a key role in organizing the large-scale magnetic field. However, in this work, the large-sale dynamo was excited regardless 
the presence of the stable layer. Then, a question naturally arises from our result ``what is the crucial 
factor which makes a difference between our work and earlier studies ?".  

To answer this question, we examine, in detail, the temporal evolution of the magnetic energy in Figure 3. The red line shows the 
volume-averaged energy of the total magnetic field $\epsilon_{\rm M,tot} \equiv \langle {\bm B}^2\rangle_{\rm v}/2\mu_0$ and the purple line is 
that of the mean-magnetic component defined by $\epsilon_{\rm M, mean} \equiv \langle {\bm B}\rangle_{\rm v}^2/2\mu_0$ for 
the model A. The blue and cyan lines are those for the model B. 

Two saturation phases appear during the evolution of the magnetic energy. The first saturation phase is shaded by dark gray 
and the second one is by light gray. This would be strongly related to the evolution of the mean magnetic component. In the first 
saturation phase, the magnetic energy of the mean component is $\mathcal O$(100) times smaller than that of the turbulent component. 
The small-scale turbulent dynamo is thus supposed to be dominated in this phase. The large-scale component evolves slowly after the first 
saturation phase over the period of the time $\sim 100 \tau_{\rm cv}$, and finally acquires a comparable strength to the turbulent field 
at around $t = 250 \tau_{\rm cv}$. 

This can be confirmed from the evolution of the spatial structure of the magnetic field. Figure 4 shows the distribution of the magnetic 
energy in the horizontal plane at the middle of the convection zone when (a) $t=100\tau_{\rm cv}$ (first saturation phase) and 
(b) $t = 330\tau_{\rm cv}$ (second saturation phase) for the model A. The filamentary small-scale magnetic field prevails in the first 
saturation phase. It inversely cascades to the larger scale as time passes, and builds up the large-scale magnetic structure in the 
second saturation phase. The magnetic structure of the model B evolves in the same manner as that of the model A. 

The evolution of the large-scale magnetic field seems to be characterized by magnetic diffusion time, which is evaluated as 
$\tau_{\rm diff} \equiv d^2/\eta_0 \simeq 250\tau_{\rm cv}$ in our models. This indicates that the simulation should be evolved for a sufficiently 
long time comparable to the magnetic diffusion time to build up a significant large-scale magnetic field. 

The resistively-dominated slow saturation of the large-scale magnetic field has been seen in the simulation of forced MHD turbulence with closed boundaries 
due to the magnetic helicity conservation (\cite{brandenburg01}). Although we have used vertical field (open) boundary condition that allows 
magnetic helicity fluxes out of the domain, it does not necessarily ensure that such fluxes are large enough to facilitate the evolution of the 
large-scale magnetic field. Even in the large-scale dynamos with open boundaries, the slow saturation has been observed (c.f., \cite{brandenburg+05,kapyla+08}). 

In \citet{favier+13}, the integration time of their simulations is substantially shorter than the magnetic diffusion time. 
%they evolved the simulation for a time substantially less than the magnetic diffusion time. 
It would be possible that the absence of large-scale dynamos in their 
study is a simple consequence of their relatively short integration time. 
To specify all the requirements for building up large-scale magnetic fields in rigidly rotating turbulent convections, we should examine long-term 
evolutions of dynamos in wider parameter range. It is however not within the scope of this work and will be the subject of our future paper. 

%%%%%%%%%%%%%%%%%%%%%%%%%%%%%%%%%%%%%%%%%%%%%%%%%%%%%%%%%%%%%%%%%%
\section{Discussion \& Summary}
%%%%%%%%%%%%%%%%%%%%%%%%%%%%%%%%%%%%%%%%%%%%%%%%%%%%%%%%%%%%%%%%%%
%%%%%%%%%%%%%%%%%%%%%%%%%%%%%%%%%%%%%%%%%%%%%%%%%%%%%%%%%%%%%%%%%%%%%%%%
\begin{figure}[tbp]
\begin{center}
\includegraphics[width=73mm]{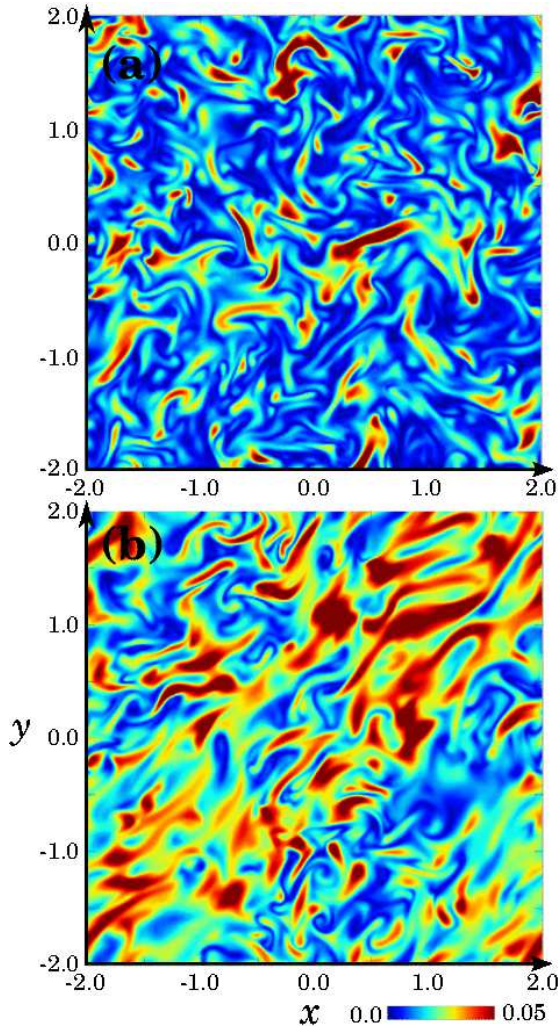}
\end{center}
\caption{Distribution of the magnetic energy on the horizontal plane at the middle of the convection zone when (a) $t=100\tau_{\rm cv}$ 
(first saturation phase) and (b) $t=330\tau_{\rm cv}$ (second saturation phase) for the model A.}
\label{fig4}
\end{figure}
%%%%%%%%%%%%%%%%%%%%%%%%%%%%%%%%%%%%%%%%%%%%%%%%%%%%%%%%%%%%%%%%%%%%%%%%
We performed rigidly rotating convective dynamo simulations in the local Cartesian geometry. 
By comparing two models with and without stably stratified layers, their effect on a large-scale dynamo was studied. 
We for the first time successfully simulated an oscillatory large-scale dynamo in the local system without the stable layer, 
whereas it was not found in the similar earlier studies (\cite{cattaneo+06,favier+13}). The absence of large-scale dynamos 
in earlier studies might be a simple consequence of their relatively short integration time. For the excitation of the large-scale 
dynamo, we should evolve the simulation for a sufficiently long time because the large-scale magnetic component is gradually 
built up in an order of the ohmic diffusion time. 

The spatiotemporal evolution of the magnetic field was similar in two models. The large-scale magnetic component was the strongest at 
around the middle of the convection layer and propagated from there to the upper and lower convection zones. According to the  mean-field 
dynamo theory, the $\alpha$-effect would be solely responsible for the large-scale dynamo in our models because the $\Omega$-effect 
is absent in the rigidly rotating system (c.f., \cite{kapyla+13}). However, the nonlinear properties of the $\alpha$-effect dynamo in the natural 
rotating convection are still veiled in mystery. We will examine quantitatively whether the $\alpha$-effect dynamo can reproduce the 
spatiotemporal evolution of large-scale magnetic fields observed at the nonlinear saturated phase in our simulations in a subsequent paper. 

An intriguing finding was the difference in the oscillation period of the large-scale magnetic field between two models. The magnetic cycle 
was about three-times longer in the model with the stable layer than without the stable layer, although the properties of the convective motion 
was similar in two models. This suggests that the stably stratified layer rather impedes the cyclic variation of large-scale magnetic fields. 
One possible cause making a difference in the cycle period is the ejection process of the magnetic helicity, which is known to affect nonlinear 
properties of dynamos (c.f., \cite{blackman+00}). 

In the model without stable layers, the magnetic helicity can be ejected from the system via advective transport processes 
because the top open boundary is placed just above the convection zone. In contrast, in the model with the top stable layer, the magnetic 
field must be transported throughout the stable layer for the loss of the magnetic helicity. The relatively slow ohmic diffusion 
dominates the transport process there. We thus speculate that 
the smaller magnetic helicity flux dominated by the slower ohmic diffusion 
process is responsible for the longer cycle period in the model with stable layers. 

The dynamo number and thus frequency of excited dynamo mode might be different between two models. 
This is because the mode with longer wavelength can be allowed in the system with conducting stable layers above and below the 
convection zone. This might be an another possibility to explain the cycle period difference (c.f., \cite{radler+87,rudiger+03}). 
In any case, further simulations with varying the thickness of the stable layers and the resistivity are necessary to %pin down 
elucidate the cause and will be a target of our future work. 

\bigskip
We acknowledge the anonymous referee for constructive comments. 
Computations were carried on XC30 at NAOJ, and K-Computer at RIKEN. This work was supported by 
JSPS KAKENHI Grant number 24740125 and the joint research project 
of the Institute of Laser Engineering, Osaka University.
%%%%%%%%%%%%%%%%%%%%%%%%%%%%%%%%%%%%%%%%%%%%%%%%%%%%%%%%%%%%%%%%%%
%%%%%%%%%%%%%%%%%%%%%%%----------------- End of Body -----------------%%%%%%%%%%%%%%%%%%%%%%%%
%%%%%%%%%%%%%%%%%%%%%%%%%%%%%%%%%%%%%%%%%%%%%%%%%%%%%%%%%%%%%%%%%%
%\bibliographystyle{apj} % for ApJ
%\bibliography{ms}

\end{document}